\documentclass[runningheads]{llncs}
\usepackage{graphicx}
\usepackage{xcolor}
\usepackage{hyperref}
\usepackage{amsfonts}
\usepackage{array,booktabs}
\newcolumntype{P}[1]{>{\centering\arraybackslash}p{#1}}
\usepackage{tikz}
\usepackage{multirow}
\usepackage{svg}
\usepackage{eucal}
\usepackage{lipsum}

\usepackage{amsmath}
\newcommand{\norm}[1]{\left\lVert#1\right\rVert}
\newcommand{\new}[1]{{#1}}

\begin{document}
%
% \title{MaskGAN: Mask-guided CycleGAN for MRI-CT Synthesis with Structural Consistency}
\title{Structure-Preserving Synthesis: MaskGAN for Unpaired MR-CT Translation\thanks{Acknowledgement: This study was supported by Channel 7 Children's Research Foundation of South Australia Incorporated (CRF).}}
% \author{Anonymous MICCAI submission}
\author{Minh Hieu Phan\inst{1}\orcidID{0000-0003-3861-0296} \and
Zhibin Liao\inst{1}\orcidID{0000-0001-9965-4511} \and
Johan W. Verjans\inst{1}\orcidID{0000-0002-8336-6774} \and
Minh-Son To\inst{2}\orcidID{0000-0002-8060-6218}}

% \authorrunning{Anonymous}
\institute{Australian Institute for Machine Learning, University of Adelaide \and
Flinders Health and Medical Research Institute, Flinders University
}

\maketitle              % typeset the header of the contribution
\begin{abstract}
   Medical image synthesis is a challenging task due to the scarcity of paired data. Several methods have applied CycleGAN to leverage unpaired data, but they often generate inaccurate mappings that shift the anatomy. This problem is further exacerbated when the images from the source and target modalities are heavily misaligned. Recently, current methods have aimed to address this issue by incorporating a supplementary segmentation network. Unfortunately, this strategy requires costly and time-consuming pixel-level annotations. To overcome this problem, this paper proposes MaskGAN, a novel and cost-effective framework that enforces structural consistency by utilizing automatically extracted coarse masks. Our approach employs a mask generator to outline anatomical structures and a content generator to synthesize CT contents that align with these structures. Extensive experiments demonstrate that MaskGAN outperforms state-of-the-art synthesis methods on a challenging pediatric dataset, where MR and CT scans are heavily misaligned due to rapid growth in children. Specifically, MaskGAN excels in preserving anatomical structures without the need for expert annotations. The code for this paper can be found at \href{https://github.com/HieuPhan33/MaskGAN}{https://github.com/HieuPhan33/MaskGAN}.

\end{abstract}
\section{Introduction}
Magnetic resonance imaging (MRI) and computed tomography (CT) are two commonly used cross-sectional medical imaging techniques. %MRI and CT use different physical principles for producing tissue contrast, and they are often used in tandem to provide complementary information. 
MRI and CT produce different tissue contrast and are often used in tandem to provide complementary information. 
While MRI is useful for visualizing soft tissues (e.g. muscle, fat), %and neural elements), 
CT is superior for visualizing bony structures. 
Some medical procedures, such as radiotherapy for brain tumors, craniosynostosis, and spinal surgery, typically require both MRI and CT for planning. 
%However, obtaining and subsequently co-registering both scans add extra cost.
Unfortunately, CT imaging exposes patients to ionizing radiation, which can damage DNA and increase cancer risk~\cite{richardson2015risk}, especially in children and adolescents.
Given these issues, there are clear advantages for synthesizing anatomically accurate CT data from MRI.

\begin{figure}[!t]
    \centering
\includegraphics[width=0.85\linewidth]{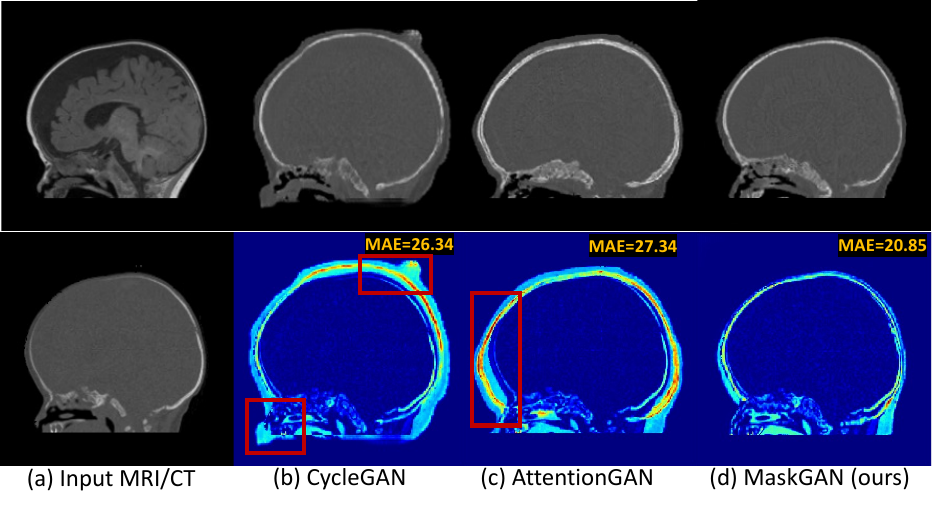}
%\vspace{-0.25em}
    \caption{Visual results (Row 1) and the error map (Row 2) between the ground-truth and synthetic CT on pediatric dataset. (a) Input MRI and the paired CT. (b) CycleGAN~\cite{zhu2017unpaired} fails to preserve the smooth anatomy of the MRI. (c) AttentionGAN~\cite{tang2021attentiongan} inflates the head area in the synthetic CT, which is inconsistent with the original MRI. %Quantitative evaluations in structural similarity index - SSIM (higher is better) and mean absolute error - MAE (lower is better) are shown in orange.
    Quantitative evaluations in MAE (lower is better) are shown in yellow.
    }
    \label{fig:teaser}
    %\vspace{-1.5em}
\end{figure}

Most synthesis methods adopt supervised learning paradigms and train generative models to synthesize CT~\cite{emami2021sa,liu2021ct,dalmaz2022resvit,zhang2022map,armanious2020medgan}.
Despite the superior performance, supervised methods require a large amount of paired data, which is prohibitively expensive to acquire. 
Several unsupervised MRI-to-CT synthesis methods~\cite{yang2020unsupervised,liu2021ct,ge2019unpaired}, leverage CycleGAN with cycle consistency supervision to %\son{(CycleGAN is not the only unsupervised method)}
eliminate the need for paired data. Unfortunately, the performance of unsupervised CT synthesis methods~\cite{yang2020unsupervised,yang2018unpaired,ge2019unpaired} is inferior to supervised counterparts.  %Specifically, they apply CycleGAN~\cite{zhu2017unpaired}, which is a state-of-the-art model for unpaired image translation. 
%While CycleGAN generates realistic images, 
%Several methods apply CycleGAN with cycle consistency loss to leverage unpaired data. However, 
Due to the lack of direct constraints on the synthetic outputs, CycleGAN~\cite{zhu2017unpaired} struggles to preserve the anatomical structure when synthesizing CT images, as shown in Fig.~\ref{fig:teaser}(b). \new{The structural distortion in synthetic results exacerbates when data from the two modalities are heavily misaligned, which usually occurs in pediatric scanning due to the rapid growth in children.}   %shows that without CycleGAN wrongfully preserves the low-intensity region corresponding to fluid at the top of the head when synthesizing CT. 
%In CycleGAN, the cycle consistency loss indirectly imposes the structural similarity between the input and the synthesized images. This often generates a mismatch in anatomical structures in the synthesized results. 

Recent unsupervised methods impose structural constraints on the synthesized CT through pixel-wise or shape-wise consistency. Pixel-wise consistency methods \cite{yang2020unsupervised,yang2018unpaired,matsuo2022unsupervised} capture and align pixel-wise correlations between MRI and synthesized CT. %capture the image structures by computing the correlations between pixels, encouraging the synthesized CT to have a similar structure as the input MRI. 
However, enforcing pixel-wise consistency may introduce undesirable artifacts in the synthetic results.
% as the real CT and MRI can have different spatial and noise statistics.
This problem is particularly relevant in brain scanning, where both the pixel-wise correlation and noise statistics in MR and CT images are different, as a direct consequence of the signal acquisition technique. 
%, \son{and noise statistics as a direct consequence of the signal acquisition technique},. %However, pixel-wise structures in the real CT differ from the MRI in certain areas, such as the brain tissue. Thus, this approach may create undesired artifacts in the synthetic CT.  %However, pixel-wise structures in the real CT can be different from the MRI in certain areas. For example, brain structures are smooth in the CT, while having complex textures in MRI. Thus, imposing MRI-like pixel-wise consistency may create undesired artifacts in the synthetic CT. %Instead of constraining pixel-wise correlations, 
The alternative shape-wise consistency methods~\cite{emami2021sa,ge2019unpaired,zhou2021anatomy}  
aim to preserve the shapes of major body parts in the synthetic image. Notably, shape-CycleGAN~\cite{ge2019unpaired} segments synthesized CT and enforces consistency with the ground-truth MRI segmentation. %enforces the segmentation to be consistent with the ground-truth MRI segmentation. %Then they minimize the discrepancy between the extracted segmentation and the ground-truth segmentation of the input MRI. 
However, these methods rely on segmentation annotations, which are time-consuming, labor-intensive, and require expert radiological annotators. %They also train additional segmentation extractor networks, which leads to computational overhead. 
A recent natural image synthesis approach, called AttentionGAN~\cite{tang2021attentiongan}, learns attention masks to identify discriminative structures. %It then synthesizes only the foreground structures. 
AttentionGAN implicitly learns prominent structures in the image without using the ground-truth shape. %In contrast to previous shape-aware methods, AttentionGAN only uses the coarse foreground masks as a hint to roughly locate the foreground structures. 
%AttentionGAN indirectly learns the foreground masks by minimizing the standard adversarial loss on the final synthetic outputs.  %AttentionGAN implicitly enforces the shape consistency when synthesizing images. 
%Despite not requiring ground-truth masks, 
Unfortunately, the lack of explicit mask supervision can lead to imprecise attention masks and, in turn, produce inaccurate mappings of the anatomy, as shown in Fig.~\ref{fig:teaser}(c).

 % \vspace{-0.5em}
\begin{table}[!h]
\centering
\renewcommand{\arraystretch}{1.05}
\setlength\tabcolsep{10.00pt}
\caption{Comparisons of different shape-aware image synthesis.}
\label{tab:compare}
\begin{tabular}{cccc}\toprule
\textbf{Method} &  \begin{tabular}{@{}c@{}}Mask \\ Supervision\end{tabular} & \begin{tabular}{@{}c@{}}Human\\ Annotation\end{tabular} & \begin{tabular}{@{}c@{}}Structural\\ Consistency\end{tabular}  \\ \midrule
Shape-cycleGAN~\cite{ge2019unpaired} & Precise mask & Yes & Yes \\
AttentionGAN~\cite{tang2021attentiongan} & Not required & No & No \\
\textbf{MaskGAN (Ours)} & \textbf{Coarse mask} & \textbf{No} & \textbf{Yes} \\ \bottomrule
\end{tabular}% 
%\vspace{-1em}
\end{table}

% \begin{figure}[!h]
%     \centering
%     \includegraphics[width=0.8\linewidth]{figures/intro.png}
%     \caption{Visual results of different methods. Column (c) and (f) show the foreground attention masks learned by AttentionGAN (AG)~\cite{tang2021attentiongan} and our MaskGAN. Inconsistent mapping problems are pronounced in the results of previous works.}
%     \label{fig:teaser}
% \end{figure}
%\vspace{-1em}

In this paper, we propose \textbf{MaskGAN}, a novel unsupervised MRI-to-CT synthesis method, that preserves the anatomy under the explicit supervision of coarse masks without using costly manual annotations. %and a new cycle shape consistency loss. 
Unlike segmentation-based methods~\cite{ge2019unpaired,zhang2018translating},
MaskGAN bypasses the need for precise annotations, replacing them with standard (unsupervised) image processing techniques, which can produce coarse anatomical masks. 
Such masks, although imperfect, provide sufficient cues for MaskGAN to capture anatomical outlines and produce structurally consistent images. Table~\ref{tab:compare} highlights our differences compared with previous shape-aware methods~\cite{ge2019unpaired,tang2021attentiongan}.
%This contrasts with previous methods, as shown in  Table~\ref{tab:compare}.
%our approach bypasses the need for precise human annotations by using automatically extracted coarse masks. Table~\ref{tab:compare} gives an overview of our method in contrast to other shape-aware methods. Standing out from recent works, MaskGAN preserves the anatomy without relying on human annotation. 
%we automatically extract the coarse ground-truth masks using image processing algorithms without relying on expensive human-annotated segmentation masks.  %We introduce a robust image processing technique based on the connected component analysis algorithm to extract the background masks from CT and MR scans. 
%Additionally, we propose a cycle shape consistency (CSC) loss to encourage the image to have a consistent foreground mask during translation. 
Our major contributions are summarized as follows. \textbf{1)} We introduce \textbf{MaskGAN}, a novel unsupervised MRI-to-CT synthesis method. MaskGAN is the first framework that maintains shape consistency without relying on human-annotated segmentation.  \textbf{2)} We present two new structural supervisions to enforce consistent extraction of anatomical structures across MRI and CT domains. \textbf{3)} Extensive experiments show that our method outperforms state-of-the-art methods by using automatically extracted coarse masks to effectively enhance structural consistency.

\section{Proposed Method}
%To resolve the inconsistent mappings, recent methods impose shape consistency between MR and synthetic CT by training an additional segmentation network~\cite{ge2019unpaired,emami2021sa}. However, they require a precise segmentation annotation, which is costly and time-consuming.
%Existing methods~\cite{zhang2022map,armanious2020medgan} employ CycleGAN~\cite{zhu2017unpaired} using the cycle consistency loss to tackle the unsupervised CT synthesis problem. However, this loss objective does not impose the structural consistency between the MRI and CT domains, leading to spurious mappings from MRI's background to CT's foreground. 
%A recent image synthesis method, called AttentionGAN~\cite{tang2021attentiongan}, implicitly learns a coarse foreground attention mask to synthesize discriminative foreground structures while leaving the background intact. However, the attention mask is learned indirectly via adversarial loss without direct mask supervision. Thus, the quality of learned masks might be poor.

\begin{figure}[!h]
    \centering
% \includegraphics[width=0.99\linewidth]{figures/maskGAN.png}
% \centerline{\includesvg[inkscapelatex=false,width=0.98\textwidth]{figures/maskgan_v2.svg}}
\includegraphics[width=1.0\textwidth]{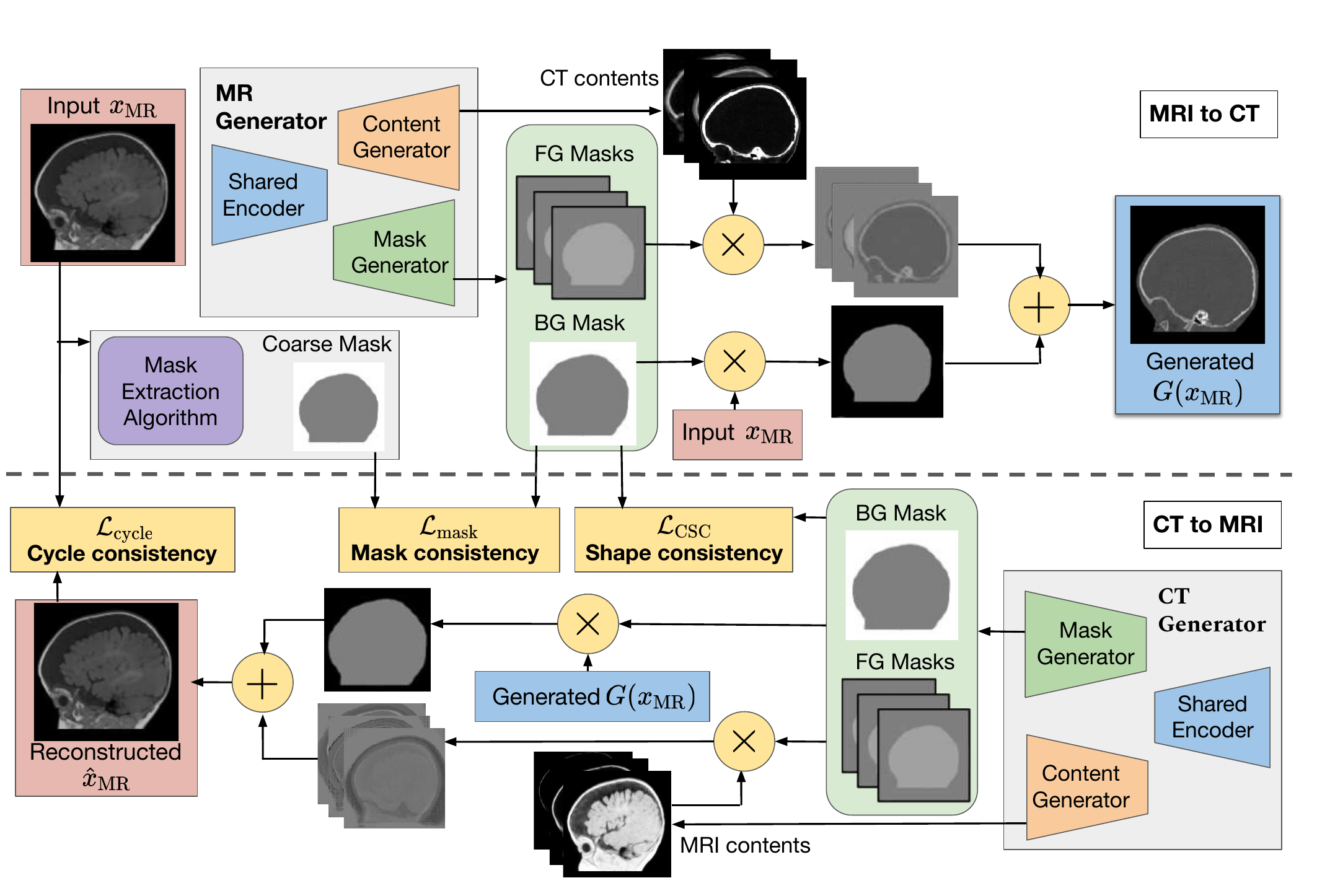}
\caption{Overview of our proposed \textbf{MaskGAN}. 
First, we automatically extract coarse masks from the input $x_\textrm{MR}$. 
MaskGAN then learns via two new objectives in addition to the standard CycleGAN loss.
The mask loss $\mathcal{L}_{\text{mask}}$ minimizes the L1 distance between the extracted background mask and the coarse mask, ensuring accurate anatomical structure generation. 
The cycle shape consistency (CSC) loss $\mathcal{L}_{\text{CSC}}$ minimizes the L1 distance between the masks learned by the MRI and CT generators, promoting consistent anatomy segmentation across domains.}
%}
\label{fig:maskGAN}
%\vspace{-0.5em}
\end{figure}

%%%%%%%%%%%%%%%%
% Why MaskGAN does not need precise segmentation map?
%%%%%%%%%%%
%\subsection{MaskGAN architecture}
%The proposed MaskGAN learns the coarse foreground and background masks, outlining the shape of prominent anatomical structures. Our emphasis is on synthesizing CT contents in the segmented foreground structures, preserving the background untouched. Fig.~\ref{fig:maskGAN} shows the MaskGAN's architecture.
In this section, we first introduce the \textbf{MaskGAN} architecture, shown in Fig.~\ref{fig:maskGAN}, and then describe the three supervision losses we use for optimization.

%Following an unsupervised learning paradigm, the proposed method learns to synthesize CT from MRI using two sets of \textit{unpaired} MR and CT images
\subsection{MaskGAN architecture} 
The network comprises two generators, each learning an MRI-CT and a CT-MRI translation. Our generator design has two branches, one for generating masks and the other for synthesizing the content in the masked regions. The mask branch learns $N$ attention masks $A_i$, where the first $N-1$ masks capture foreground (FG) structures and the last mask represents the background (BG). The content branch synthesizes $N-1$ outputs for the foreground structures, denoted as $C$. Each output, $C_i$, represents the synthetic content for the corresponding foreground region that is masked by the attention mask $A_i$. % $i$-th synthetic content $C_i$ is for the $i$-th foreground region masked by $A_i$. 

Intuitively, each channel $A_i$ in the mask tensor $A$ focuses on different anatomical structures in the medical image. For instance, one channel emphasizes on synthesizing the skull, while another focuses on the brain tissue. %Here, the attention mask tensor $A$ has a shape of $N \times H \times W$, where $N$ is the total number of channels, $H \times W$ denote the height and width of the input and output image. 
The last channel $A_N$ in A corresponds to the background and is applied to the original input to preserve the background contents. The final output is the sum of masked foreground contents and masked background input. Formally, the synthetic CT output generated from the input MRI $x$ is defined as
\begin{equation}
    O_{\text{CT}} = G_{\text{CT}}(x) = A_{\text{CT}}^N x + \sum_{i=1}^{N-1} A_{\text{CT}}^i C_{\text{CT}}^i.
\end{equation}
The synthetic MRI output from the CT scan $y$ is defined similarly based on the attention masks and the contents from the MR generator. %Fig.~\ref{fig:maskGAN} shows the overview of our MaskGAN with $N=3$ attention masks. 
The proposed network is trained using three training objectives described in the next sections.%: the standard CycleGAN loss, our proposed mask supervision, and shape consistency supervision.

\subsection{CycleGAN supervision}
\label{method:cyclegan}
The two generators, $G_{MR}$ and $G_{CT}$, map images from MRI domain ($X$) and CT domain ($Y$), respectively. Two discriminators, $D_{MR}$ and $D_{CT}$, are used to distinguish real from fake images in the MRI and CT domains. The adversarial loss for training the generators to produce synthetic CT images is defined as
\begin{equation}
    \resizebox{0.99\hsize}{!}{    
    $\mathcal{L}_{\text{CT}}(G_{\text{MR}}, D_{\text{CT}},x,y) = \mathbb{E}_{y \sim p_{\text{data}}(y)} \big[ \log D_{\text{CT}}(y) \big] +  \mathbb{E}_{x \sim p_{\text{data}}(x)} \big[ \log (1- D_{\text{CT}}(G_{\text{MR}}(x))) \big]$.
    }
\end{equation}
The adversarial loss $\mathcal{L}_{\text{MR}}$ for generating MRI images is defined in a similar manner.
% \begin{equation}
%     \resizebox{0.99\hsize}{!}{   
%     $L_{\text{adv}}(G_{\text{CT}}, D_{\text{MRI}},y,x) = \mathbb{E}_{x \sim p_{\text{data}}}(x) \big[ \log D_{\text{MRI}}(x) \big] +  \mathbb{E}_{x \sim p_{\text{data}}}(x) \big[ \log (1- D_{\text{CT}}(G_{\text{MR}}(x))) \big]$.
%     }
% \end{equation}
For unsupervised training, CycleGAN imposes the cycle consistency loss, which is formulated as follows
\begin{equation}
    \mathcal{L}_{\text{cycle}} = \mathbb{E}_{x \sim p_{\text{data}}}(x) \norm{x - G_{CT}(G_{MR}(x))} + \mathbb{E}_{y \sim p_{\text{data}}}(y) \norm{y - G_{MR}(G_{CT}(y))}.
\end{equation}
The CycleGAN's objective $\mathcal{L}_{\text{GAN}}$ is the combination of adversarial and cycle consistency loss.%: $L_{\text{GAN}} = \lambda L_{\text{cycle}} + L_{\text{MRI}} + L_{\text{CT}}$, 
% \begin{equation}
%     L_{\text{GAN}} = \lambda L_{\text{cycle}} + L_{\text{MRI}} + L_{\text{CT}},
% \end{equation}
%where $\lambda$ is a coefficient controlling the strength of cycle consistency loss.

%%%
\subsection{Mask and cycle shape consistency supervision}
\label{method:mask}
\textbf{Mask loss.} To reduce spurious mappings in the background regions, MaskGAN explicitly guides the mask generator to differentiate the foreground objects from the background using mask supervision. We extract the coarse mask $B$ using basic image processing operations.
Specifically, we design a simple but robust algorithm that works on both MRI and CT scans, with a binarization stage followed by a refinement step. 
\new{In the binarization stage, we normalize the intensity to the range [0, 1] and apply a binary threshold of 0.1, selected based on histogram inspection, to separate the foreground from the background. 
In the post-processing stage, we refine the binary image using morphological operations, specifically employing a binary opening operation to remove small artifacts. We perform connected component analysis~\cite{samet1988efficient} and keep the largest component as the foreground. Column 6 in Fig.~\ref{fig:vis} shows examples of extracted masks.} %Connected component analysis~\cite{samet1988efficient}, a well-established algorithm for blob extraction, is applied to capture the largest component in the binary image as the foreground structure. 

We introduce a novel mask supervision loss that penalizes the difference between the background mask $A_N$ learned from the input image and the ground-truth background mask $B$ in both MRI and CT domains. The mask loss for the attention generators is formulated as
\begin{equation}
    \mathcal{L}_{\text{mask}} = \mathbb{E}_{x \sim p_{\text{data}}(x)} \norm{A^{\text{MR}}_N - B^{\text{MR}}_N}_1 + \mathbb{E}_{y \sim p_{\text{data}}(y)} \norm{A_N^{\text{CT}} - B_N^{\text{CT}}}_1.
\end{equation}

%\textit{Discussion.} Previous segmentation-based methods~\cite{ge2019unpaired,zhang2018translating} only have a single content generator and enforce shape consistency on the \textit{synthetic contents} from the generator. In contrast, we decouple the learning into two generators and only impose consistency on the \textit{mask} produced by the mask generator, while giving the content generator more flexibility. This flexibility allows the content generator to adjust content when the learned mask is imperfect. Disentangling the generator into two branches for learning masks and contents separately makes our MaskGAN more robust to error-prone segmentation masks. 

\textit{Discussion.} Previous shape-aware methods~\cite{ge2019unpaired,zhang2018translating} use a pre-trained U-Net~\cite{ronneberger2015u} segmentation network to enforce shape consistency on the generator. U-Net is pre-trained in a separate stage and frozen when the generator is trained. Hence, any errors produced by the segmentation network cannot be corrected. In contrast, we jointly train the shape extractor, \textit{i.e.}, the mask generator, and the content generator end-to-end. Besides mask loss $\mathcal{L}_{\text{mask}}$, the mask generator also receives supervision from an adversarial loss $\mathcal{L}_{\text{GAN}}$ to adjust the extracted shape and optimize the final synthetic results. %This strategy allows the both generators to communicate and potentially adjust the extracted shape to optimize the final synthetic results. 
Moreover, in contrast to previous methods that train a separate shape extractor, our MaskGAN uses a shared encoder for mask and content generators, as illustrated in Fig.~\ref{fig:maskGAN}. Our design embeds the extracted shape knowledge into the content generator, thus improving the structural consistency of the synthetic contents.

\textbf{Cycle shape consistency loss.}
Spurious mappings can occur when the anatomy is shifted during translation. To preserve structural consistency across domains, we introduce the cycle shape consistency (CSC) loss as our secondary contribution. 
%An image should have the same anatomical structures in MRI and CT domains. To preserve the structural consistency when translating across domains, we propose a cycle shape consistency (CSC) loss. 
Our loss penalizes the discrepancy between the background attention mask $A^{\text{MR}}_N$ learned from the input MRI image and the mask $\tilde{A}^{\text{CT}}_N$ learned from synthetic CT. Enforcing consistency in both domains, we formulate the shape consistency loss as
\begin{equation}
    \mathcal{L}_{\text{shape}} =  \mathbb{E}_{x \sim p_{\text{data}}(x)} \norm{A_N^{\text{MR}} - \tilde{A}_N^{\text{CT}}}_1 + \mathbb{E}_{y \sim p_{\text{data}}(y)} \norm{A_N^{\text{CT}} - \tilde{A}_N^{\text{MR}}}_1.
\end{equation}
%Applying shape consistency on each individual foreground channel $A_i$ between two domains may restrict the learning flexibility of the two generators. Thus, we only apply shape consistency loss on the background mask $A_N$. 
%Our strategy allows the MRI and CT generator to focus on translating different foreground areas. 
The final loss for MaskGAN is the sum of three loss objectives weighted by the corresponding loss coefficients: $\mathcal{L} =\mathcal{L}_{\text{GAN}} + \lambda_{\text{mask}}\mathcal{L}_{\text{mask}} + \lambda_{\text{shape}}\mathcal{L}_{\text{shape}}$.
% \begin{equation}
%     \mathcal{L} =\mathcal{L}_{\text{GAN}} + \lambda_{\text{mask}}\mathcal{L}_{\text{mask}} + \lambda_{\text{shape}}\mathcal{L}_{\text{shape}}
% \end{equation}
%MRI and CT images have different contrasts in certain body areas. For example, bones have high contrast in CT, while having low values in MRI. The foreground masks in the MRI image could be different from the CT image. Therefore, we only apply shape consistency loss on the background mask $A_N$ but not on each individual foreground mask $A_i$. Our strategy allows the MRI generator and CT generator to independently learn different and diverse foreground masks.

% \textbf{Discussion on learnable attention masks.} Instead of learning the background masks, one can directly use the ground-truth background mask. However, ground-truth background masks have many artifacts since traditional image processing techniques are susceptible to noise. Thus, our MaskGAN learns the attention masks 

\section{Experimental Results}
\subsection{Experimental Settings}
\textbf{Data collection.} %Ethics approval for this study was provided by \emph{anonymized}. 
%Ethics approval for this study was granted by Anonymous. 
We collected 270 volumetric T1-weighted MRI and 267 thin-slice CT head scans with bony reconstruction performed in pediatric patients under routine scanning protocols\footnote{Ethics approval was granted by Southern Adelaide Clinical Human Research Ethics Committee}.
We targeted the age group from 6–24 months since pediatric patients are more susceptible to ionizing radiation and experience a greater cancer risk (up to 24\% increase) from radiation exposure~\cite{mathews2013cancer}. 
Furthermore, surgery for craniosynostosis, a birth defect in which the skull bones fuse too early, typically occurs during this age~\cite{zakhary2014surgical,governale2015craniosynostosis}. 
%. Geometric distortions in MR volumes were corrected using a 3D correction algorithm in the [Siemens Syngo console workstation]. All the MR volumes were N4 corrected and normalized by aligning the white matter peak identified by the fuzzy C-means. 
The scans were acquired by %MR/CT Philips and GE system	
Ingenia 3.0T MRI scanners and Philips Brilliance 64 CT scanners. We then resampled the volumetric scans to the same resolution of 1.0 $\times$ 1.0 $\times$ 1.0 $\text{mm}^3$. %The volumetric scans were resampled to the same resolution of 1.0 $\times$ 1.0 $\times$ 1.0 $\text{mm}^3$. %MRI and CT volumes were resampled to the same resolution of 1.0 $\times$ 1.0 $\times$ 1.0 $\text{mm}^3$. 
 %13 MRI-CT volumes from the same patients that were captured less than three months apart are registered using rigid registration algorithms with the Dipy Python library.%\footnote{\url{https://dipy.org/}.}.

The dataset comprises brain MR and CT volumes from 262 subjects. 13 MRI-CT volumes from the same patients that were captured less than three months apart are registered using rigid registration algorithms. 
 The dataset is divided into 249, 1 and 12 subjects for training, validating and testing set. Following~\cite{wolterink2017deep}, we conducted experiments on sagittal slices.  Each MR and CT volume consists of 180 to 200 slices, which are resized and padded to the size of $224 \times 224$. The intensity range of CT is clipped into [-1000, 2000]. All models are trained using the Adam optimizer for 100 epochs, with a learning rate of 0.0002 which linearly decays to zero over the last 50 epochs. We use a batch size of 16 and train on two NVIDIA RTX 3090 GPUs. %using PyTorch version 1.9.1. %The hyperparameters for each loss are fine-tuned to achieve optimal performance. %The Supplementary material describe the ablation studies on the hyperparameters.

\textbf{Evaluation metrics.} To provide a quantitative evaluation of methods, we compute the same standard performance metrics as in previous works~\cite{yang2020unsupervised,liu2021ct} including  mean absolute  error (MAE),  peak  signal-to-noise ratio (PSNR), and structural similarity (SSIM) between ground-truth and synthesized CT. \new{The scope of the paper centers on theoretical development; clinical evaluations such as dose calculation and treatment planning will be conducted in future work.}

\subsection{Results and Discussions}
\textbf{Comparisons with state-of-the-art.} We compare the performance of our proposed MaskGAN with existing state-of-the-art image synthesis methods, including CycleGAN~\cite{zhu2017unpaired}, AttentionGAN~\cite{tang2021attentiongan},  structure-constrained CycleGAN (sc-CycleGAN)~\cite{yang2020unsupervised} and shape-CycleGAN~\cite{ge2019unpaired}. Shape-CycleGAN requires annotated segmentation to train a separate U-Net. For a fair comparison, we implement shape-CycleGAN using our extracted coarse masks based on the authors' official code.%\footnote{\url{https://github.com/gyhandy/Unpaired_MR_to_CT_Image_Synthesis}.}. %We implement shape-CycleGAN following the official author's codebase\footnote{\url{https://github.com/gyhandy/Unpaired_MR_to_CT_Image_Synthesis}}.
%For shape-CycleGAN, following author's codebase\footnote{\href{https://github.com/gyhandy/Unpaired_MR_to_CT_Image_Synthesis}{https://github.com/gyhandy/Unpaired_MR_to_CT_Image_Synthesis}}, we first train the U-Net using our extracted coarse masks, freezes U-Net and trains the generators. %Among them, sc-CycleGAN and AttentionGAN are structural-based image synthesis methods which aim to preserve the structures of the image in the input domain. 
Note that CT-to-MRI synthesis is a secondary task supporting the primary MRI-to-CT synthesis task. As better MRI synthesis leads to improved CT synthesis, we also report the model's performance on MRI synthesis.

\begin{table}[!h]
    \caption{Quantitative comparison of different methods on the primary MRI-CT task and the secondary CT-MRI task. The results of an ablated version of our proposed MaskGAN are also reported. $\pm$ standard deviation is reported over five evaluations. The paired t-test is conducted between MaskGAN and a compared method at $p=0.05$. The improvement of MaskGAN over all compared methods is statistically significant.}
    \label{tab:sota}
    \centering
    \resizebox{\columnwidth}{!}{%
    \begin{tabular}{@{}l|ccc||ccc@{}}
    \toprule \multirow{ 2}{*}{\textbf{Methods}} & \multicolumn{3}{c}{\textbf{Primary: MRI-to-CT}} & \multicolumn{3}{c}{\textbf{Secondary: CT-to-MRI}} \\
         & MAE $\downarrow$ & PSNR $\uparrow$ & SSIM (\%) $\uparrow$ & MAE $\downarrow$ & PSNR $\uparrow$ & SSIM (\%) $\uparrow$  \\\midrule
         CycleGAN~\cite{zhu2017unpaired} & {32.12 $\pm$ 0.31} &{31.57 $\pm$ 0.12} &{46.17 $\pm$ 0.20 } &{34.21 $\pm$ 0.33} &{29.88 $\pm$ 0.24} &{45.73 $\pm$ 0.17} \\
    % MedGAN~\cite{armanious2020medgan} & { 32.54 & {32.57 & {52.67 & {30.87 & {31.45 & {50.57\\
AttentionGAN~\cite{tang2021attentiongan} & {28.25 $\pm$ 0.25} &{32.88 $\pm$ 0.09} &{53.57 $\pm$ 0.15} &{30.47 $\pm$ 0.22} &{30.15 $\pm$ 0.10} &{50.66 $\pm$ 0.14} \\
         sc-CycleGAN~\cite{yang2020unsupervised} & {24.55 $\pm$ 0.24} &{32.97 $\pm$ 0.07} &{57.08 $\pm$ 0.11} &{26.13 $\pm$ 0.15} &{31.22 $\pm$ 0.07 } &{54.14 $\pm$ 0.10} \\
         shape-CycleGAN~\cite{ge2019unpaired} & {24.30 $\pm$ 0.28} &{33.14 $\pm$ 0.05} &{57.73 $\pm$ 0.13} &{25.96 $\pm$ 0.19} &{31.69 $\pm$ 0.08} &{54.88 $\pm$ 0.09} \\\midrule
        MaskGAN (w/o Shape) & { 22.78 $\pm$ 0.19} &{ 34.02 $\pm$ 0.09} &{60.19 $\pm$ 0.06} &{23.58 $\pm$ 0.23} &{32.43 $\pm$ 0.07} &{57.35 $\pm$ 0.08} \\
        %MaskGAN (w/o Mask) & {24.15 $\pm$ 0.22} &{ 33.54 $\pm$0.10} &{59.93 $\pm$ 0.11} &{23.11 $\pm$ 0.28} &{32.06 $\pm$ 0.09} &{56.88 $\pm$ 0.12} \\
         MaskGAN (Ours) & \textbf{21.56 $\pm$ 0.18}  & \textbf{34.75 $\pm$ 0.08} & \textbf{61.25 $\pm$ 0.10} & \textbf{22.77 $\pm$ 0.17} & \textbf{32.55 $\pm$ 0.06} & \textbf{58.32 $\pm$ 0.10} \\\hline
    \end{tabular}%
    }
\end{table}

\begin{figure}[!h]
    \centering
\includegraphics[width=1.0\linewidth]{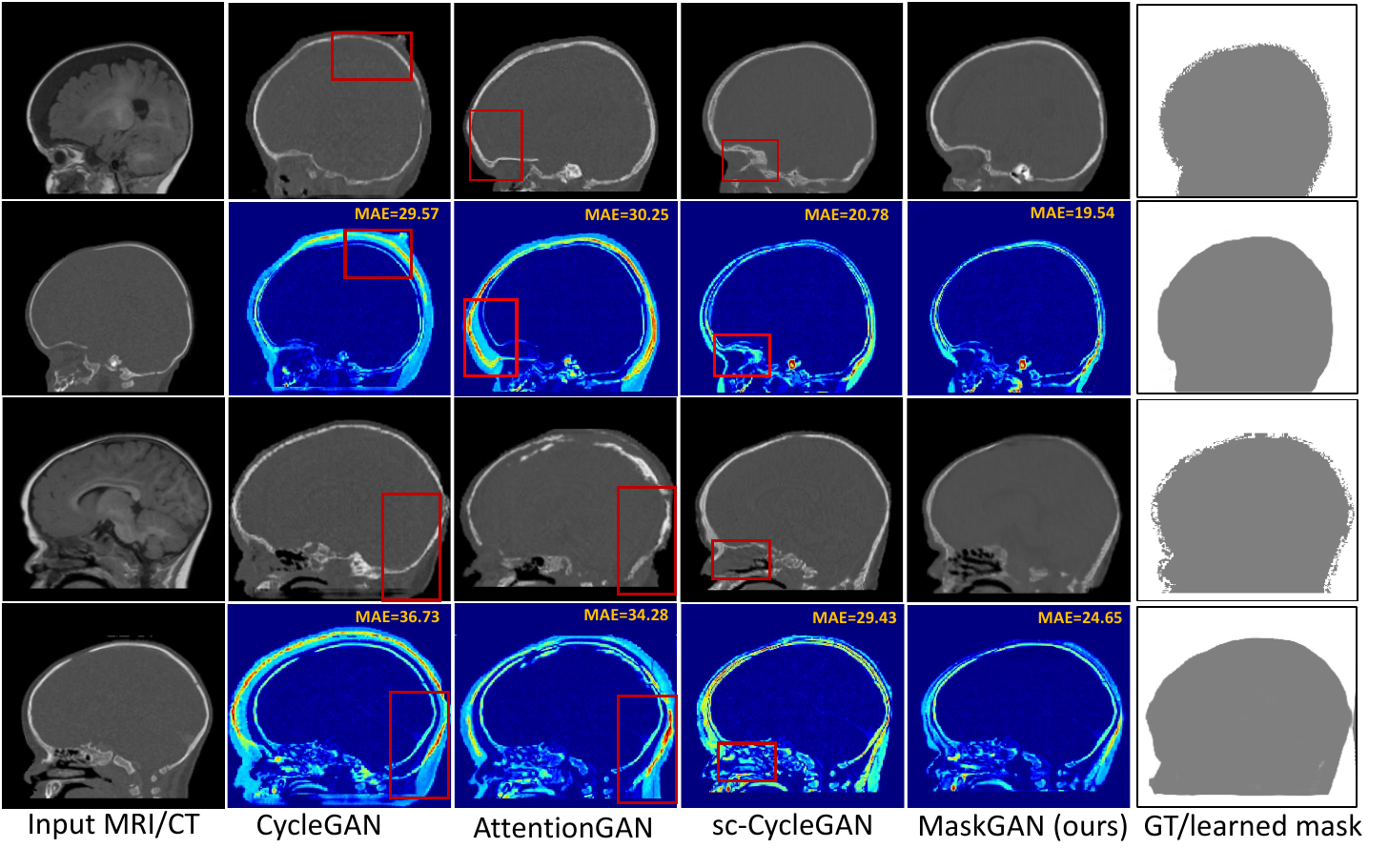}
%\vspace{-1.5em}
    \caption{Visual comparison of synthesized CT images by different methods on two samples. Column 1: Input MRI (Row 1 and 3) and the corresponding paired CT scan (Row 2 and 4). Column 2-5: Synthesized CT results (Row 1 and 3) and the corresponding error maps (Row 2 and 4). Column 6: Extracted coarse background (ground-truth) masks (Row 1 and 3) and attention masks learned by our MaskGAN (Row 2 and 4).}
    \label{fig:vis}
%\vspace{-1em}
\end{figure}

Table~\ref{tab:sota} demonstrates that our proposed MaskGAN outperforms existing methods for statistical significance of $p=0.05$ in both tasks.  %Our method drastically decreases the MAE of CycleGAN by 44.17\%. Compared with structural-based methods, the proposed method improves the SSIM of sc-CycleGAN and AttentionGAN by 13.86\% and 14.95\%, respectively. Compared with the shape-aware method, Oour MaskGAN reduces MAE of shape-CycleGAN by 21.44\%. 
The method reduces the MAE of CycleGAN and AttentionGAN by 29.07\% and 19.36\%, respectively. %and significantly improves the SSIM of AttentionGAN by 14.11\%. 
%Compared with sc-CycleGAN, which is the current state-of-the-art method in unsupervised MR-CT synthesis, our MaskGAN drastically boosts the SSIM by 7.31\%. 
Furthermore, MaskGAN outperforms shape-CycleGAN, reducing its MAE by 11.28\%. Unlike shape-CycleGAN, which underperforms when trained with coarse segmentations, our method obtains consistently higher results. Fig.~\ref{fig:vis} shows the visual results of different methods. sc-CycleGAN produces artifacts (e.g., the eye socket in the first sample and the nasal cavity in the second sample), as it preserves pixel-wise correlations. In contrast, our proposed MaskGAN preserves shape-wise consistency and produces the smoothest synthetic CT. \new{Unlike adult datasets~\cite{yang2020unsupervised,ge2019unpaired}, pediatric datasets are easily misaligned due to children's rapid growth between scans. Under this challenging setting, unpaired image synthesis can have non-optimal visual results and SSIM scores. Yet, our MaskGAN achieves the highest quality, indicating its suitability for pediatric image synthesis.} %Compared with AttentionGAN, our MaskGAN better preserves the head shape and creates the most consistent MR-to-CT mapping. %In contrast, AttentionGAN and our proposed MaskGAN preserve shape-wise consistency, producing smoother CT outputs. Compared with AttentionGAN, our MaskGAN better delineates the head shape and creates the most consistent mapping from MRI to CT.

We perform an ablation study by removing the cycle shape consistency loss (w/o Shape).
%To validate the impact of the two proposed loss objectives (mask loss and cycle shape consistency loss), two ablated versions of MaskGAN are evaluated. %They are MaskGAN trained with CSC loss only, i.e., MaskGAN (w/o CSC) with mask loss only, i.e., MaskGAN (w/o Shape). %Note that AttentionGAN~\cite{tang2021attentiongan} is MaskGAN without mask and shape-consistency losses. 
Compared with shape-CycleGAN, MaskGAN using only a mask loss significantly reduces MAE by 6.26\%. %The background masks implicitly learned by AttentionGAN can be unsatisfactory. In contrast, our proposed mask loss explicitly supervises the attention learning process using the ground-truth mask. Hence, applying our mask loss significantly improves AttentionGAN in all metrics. Our MaskGAN combines both mask loss and shape-consistency loss to ensure the structural consistency when translating between MRI and CT domains. 
The combination of both mask and cycle shape consistency losses results in the largest improvement, demonstrating the complementary contributions of our two losses.%, further decreasing MAE of MaskGAN (w/o Shape) by 5.36\% and 3.44\% on primary and secondary tasks, respectively. 

\begin{figure}[!h]
     \centering
\includegraphics[width=0.92\linewidth]{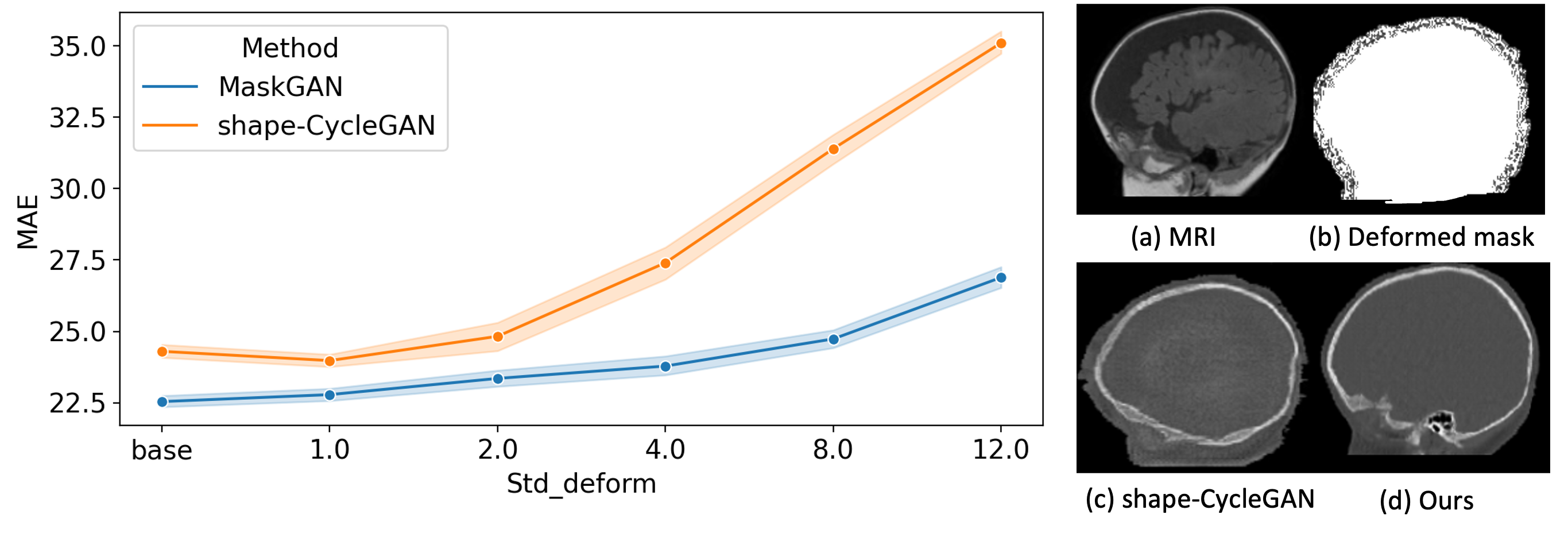}
%\vspace{-1.0em}
     \caption{\textit{Left}: MAE of the two shape-aware methods using deformed  masks. \textit{Right}: Qualitative results of shape-CycleGAN (c) and our MaskGAN (d) when training using coarse masks deformed by the standard deviation of 2.0.}
     \label{fig:deform}
     %\vspace{-1.5em}
 \end{figure}

\textbf{Robustness to error-prone coarse masks.}
%We investigate the robustness of MaskGAN to noisy ground-truth masks and 
We compare the performance of our approach with shape-CycleGAN~\cite{ge2019unpaired} using deformed masks that simulate human errors during annotation. %To alter object shapes, we employ random elastic deformation, which is a standard data augmentation technique in medical image segmentation~\cite{ronneberger2015u}. This technique applies random displacement vectors to objects. The level of distortion is controlled by the standard deviation of the normal distribution from which the vectors are sampled.
To alter object shapes, we employ random elastic deformation, a standard data augmentation technique~\cite{ronneberger2015u} that applies random displacement vectors to objects. The level of distortion is controlled by the standard deviation of the normal distribution from which the vectors are sampled.
Fig.~\ref{fig:deform} (Left) shows MAE of the two methods under increasing levels of distortion. MAE of shape-CycleGAN drastically increases as the masks become more distorted. Fig.~\ref{fig:deform} (Right) shows that our MaskGAN (d) better preserves the anatomy. %In contrast to shape-CycleGAN, which freezes the shape extractor, our technique jointly trains the shape and the content generators. Besides mask loss, the mask generator also receives supervision from the adversarial loss to remedy the impact of error-prone segmentation. %This joint training remedies the impact of error-prone segmentation. %Besdies mask loss, the mask generator also receives supervision from adversarial loss. Hence, the generator can learn to adjust the extracted shape to optimize the final output. %Jointly training shape and content generators as in our MaskGAN can remedy the impact of error-prone segmentation. %This highlights the robustness of MaskGAN, even when trained under error-prone masks.

\section{Conclusion}
This paper proposes MaskGAN - a novel automated framework that maintains the shape consistency of prominent anatomical structures without relying on expert annotated segmentations. Our method generates a coarse mask outlining the shape of the main anatomy and synthesizes the contents for the masked foreground region. Experimental results on a clinical dataset show that MaskGAN significantly outperforms existing methods and produces synthetic CT with more consistent mappings of anatomical structures.

\bibliographystyle{splncs04}
\bibliography{paper654}

\end{document}